\documentclass[12pt,preprint]{aastex}

\shorttitle{Rossby}
\shortauthors{Zaqarashvili et al.}

\begin{document}

\title{Dynamics of a solar prominence tornado observed by SDO/AIA on 2012 November 7-8}

\author{Irakli Mghebrishvili\altaffilmark{1}, Teimuraz V. Zaqarashvili\altaffilmark{2,1}, Vasil Kukhianidze\altaffilmark{1}, Giorgi Ramishvili\altaffilmark{1}, Bidzina Shergelashvili\altaffilmark{1,4}, Astrid Veronig\altaffilmark{3} and Stefaan Poedts\altaffilmark{4}
}

\altaffiltext{1}{Abastumani Astrophysical Observatory at Ilia State University, University St. 2, Tbilisi, Georgia}
\altaffiltext{2}{Space Research Institute, Austrian Academy of Sciences, Schmiedlstrasse 6, 8042 Graz, Austria, Email: teimuraz.zaqarashvili@oeaw.ac.at}
\altaffiltext{3}{IGAM-Kanzelh\"ohe Observatory, Institute of Physics, University of Graz, Universit\"atsplatz 5, 8010 Graz, Austria}
\altaffiltext{4}{Centre for Mathematical Plasma Astrophysics, Celestijnenlaan 200B, 3001, Leuven, Belgium}

\begin{abstract}

We study the detailed dynamics of a solar prominence tornado using time series of 171, 304, 193 and 211 {\AA} spectral lines obtained by \emph{Solar Dynamics Observatory}/ Atmospheric Imaging Assembly during 2012 November 7-8.
The tornado first appeared at 08:00 UT, November 07, near the surface, gradually rose upwards with the mean speed of $\sim$ 1.5 km s$^{-1}$ and persisted over 30 hr. Time-distance plots show two patterns of quasi-periodic transverse displacements of the tornado axis with periods of 40 and 50 minute at different phases of
the tornado evolution. The first pattern occurred during the rising phase and can be explained by the upward motion of the twisted tornado.
The second pattern occurred during the later stage of evolution when the tornado already stopped rising and could be caused either by MHD kink waves in the tornado or by the rotation of two tornado threads around a common axis. The later hypothesis is supported by the fact that the tornado sometimes showed a double structure during the quasi-periodic phase. 211 and 193 {\AA} spectral lines show a coronal cavity above the prominence/tornado, which started expansion at $\sim$ 13:00 UT and continuously rose above the solar limb. The tornado finally became unstable and erupted together with the corresponding prominence as coronal mass ejection (CME) at 15:00 UT, November 08. The final stage of the evolution of the cavity and the tornado-related prominence resembles the magnetic breakout model. On the other hand, the kink instability may destabilize the twisted tornado, and consequently prominence tornadoes can be used as precursors for CMEs.

\end{abstract}


\section{Introduction}\label{intro}

Solar giant tornadoes appear in the corona like a twisted rope or a fine screw \citep{Pettit1925}.
They are usually related to filaments/prominences which are called "tornado prominences" \citep{Zolner1869, Pettit1950}. The tornadoes are probably formed either due to the expansion of a twisted flux rope into the coronal cavity \citep{li2012,Panesar2013} or due to large vortex flows, which are frequently observed in the solar photosphere \citep{Brandt1988, Attie2009}. Recent observations also showed the existence of small scale tornadoes (magnetic tornadoes) in the chromosphere, which may provide an alternative mechanism for channeling energy from the lower into the upper solar atmosphere \citep{Wedemeyer2012}. The small scale tornadoes are probably formed by small scale photospheric vortex flows, which are frequently observed and also frequently occur in numerical simulations \citep{Stein,Bonet2008,Bonet2010,Wedemeyer2009,Steiner}.

Solar prominences are large magnetic structures confining cool and dense plasma in the hot solar corona. They may persist from a few days to several months. Prominences/filaments may undergo large-scale instabilities which may lead to their eruption. These eruptions are often associated with flares and coronal mass ejections (CMEs) \citep{Labrosse2010,Mackay2010}. Solar tornadoes, which are associated with prominences, can erupt together with the filament spine \citep{Su2012,Wedemeyer2013}. Tornadoes may play a distinct role in the supply of mass and twists to filaments \citep{Su2012}. Recently \citet{Wedemeyer2013} and \citet{Su2014} observed Doppler shifts in tornado-like prominences and showed that the tornadoes under study revealed persistent redshifts and blueshifts on the opposite sides of the tornado axis, which means that tornadoes are indeed rotating structures and the rotation is not a "vortical illusion" \citep{Panasenco2014}.

Observations show frequent small and large amplitude oscillations in solar prominences \citep{Arregui2012}. Large-amplitude oscillations are caused by Moreton waves \citep{Ramsey1966}, EIT waves \citep{Okamoto2004}, nearby subflares \citep{Jing2003, Vrsnak2007} or local flux emergence near the filaments \citep{Isobe2006}. In some cases, the oscillations are associated to the eruptive phase of a filament \citep{Isobe2006,Isobe2007}. On the other hand, small-amplitude oscillations have no particular triggering sources and they are frequently observed in solar prominence threads \citep{Oliver2002,Lin2005,Lin2007,Lin2009,Mackay2010}. As giant tornadoes are generally associated with prominences/filaments, they may also show some sort of oscillatory motions.

{Quiescent prominences are often surrounded by coronal cavities, which appear as dark semicircular or circular regions in the corona. Coronal cavities and prominences may erupt together as CMEs. Consequently, cavities are often observed in CMEs with three-part structure \citep{Illing1985,Chen1997,Dere1999}. Therefore, to study the relation between giant tornadoes and coronal cavities is an interesting problem. Giant tornadoes might be related to the destabilization of prominences/filaments and the consequent initiation of CMEs, therefore studies of their evolution is relevant for solar and space weather predictions. In this paper we present a case study of the formation, evolution and eruption of a solar giant tornado observed by Atmospheric Imaging Assembly (AIA)/\emph{Solar Dynamics Observatory (SDO)}.

\section{Observations and data analysis}

We use observational data obtained by the AIA \citep{Lemen2012} on board the \emph{SDO} \citep{Pesnell(2012)} on 2012
November 7-8. AIA provides full disk observations of the Sun in three ultraviolet continua and seven EUV narrow band channels with 1''.0 resolution and 12 s cadence. We use level 1.0 images of the 171, 304, 193 and 211 {\AA} band channels. The level 1.0 images include the bad-pixel removal, despiking and flat-fielding. The data are calibrated and analyzed using standard routines in the SolarSoft (SSW) package.

The event has been observed from 08:00 UT, 2012 November 07, to 15:30 UT, November 08. The time series shows that the tornado started to form at 08:00 UT on 2012 November 07 on the solar southwest limb and persisted over $\sim$ 30 hr. The tornado was associated with a solar prominence, which eventually erupted as a CME. Figure~\ref{fig1} shows the formation and evolution of the tornado in the 171 {\AA} filter during the whole life-time. Tornadoes are usually visible as a dark absorption structure in the 171 {\AA} filter, but they may appear as emitting structures in the 304 {\AA} filter \citep{Wedemeyer2013}. The top-left panel shows the coronal image at 15:01 UT, November 06, where the tornado was not yet seen. The white arrow indicates the place where the tornado was formed later. The top-right panel shows the tornado at 09:25 UT, November 07, in the form of a thin black thread (the corresponding image in the 304 {\AA} line is shown on upper left panel of Figure~\ref{fig7}). The middle left panel shows the image at 15:01 UT, November 07, where the tornado had already risen well above the solar surface. The tornado was already well developed at 23:37 UT, November 07 as it is seen on the middle right panel. The lower left panel displays the tornado at 10:25 UT, November 08, which shows that the tornado stayed stable over many hours. Finally, it started to become unstable and erupted together with the associated prominence at 15:00 UT, November 08, as can be observed on the lower right panel.

It is not possible to check the relation of the tornado formation to photospheric vortices because of the limb location. On the other hand, there is some evidence in movies that the tornado was rotating during its whole lifetime. It is desirable to check the rotation by spectroscopic observations, but spectroscopic observations of tornadoes are extremely rare. However, recently \citet{Su2014}, performed a dedicated EIS/\emph{Hinode} observing campaign to study the plasma motions in tornadoes.

During its evolution, the tornado was continuously increasing in height: it reached the maximal height of 50 Mm in approximately 10 hr after the first appearance i.e. at $\sim$ 18:00 UT, November 07. Therefore, the mean rising speed can be estimated as $\sim$ 1.5 km s$^{-1}$. When reaching the maximal height, the tornado kept this height over the next 10 hr.

The width of the tornado was also continuously changing with time and height. Sometimes it split into two threads, but the threads reunited again later on.
It is seen that the tornado generally was narrower near the footpoint and wider near the top. Such dependence of the width on height seems to be characteristic for tornados \citep{Su2012,Wedemeyer2013}.

\section{Results}

In order to study in detail the temporal dynamics of the tornado, we constructed time-distance plots at six different height levels above the limb. The location of these six cuts is shown on middle right panel of Figure~\ref{fig1} by white solid lines. The first cut (indicated by 1) is located at the height of $\sim$ 14 Mm and the sixth cut (indicated by 6) is located at a height of $\sim$ 44 Mm above the solar limb. The distance between each of the five lines from cut 2 to cut 6 is 2.9 Mm.

Figure~\ref{fig2} shows the time-distance diagram at the cut 1 (upper panel) and the cut 6 (lower panel) during its full life-time from 05:00 UT, 2012 November 07, to 15:00 UT, November 08. On the lower panel it is seen how the tornado appeared at the height of the cut 6 ($\sim$ 44 Mm) near 17:00 UT, November 07. There are apparent quasi-periodic transverse displacements of the axis during different intervals during its lifetime. We select and study two clear patterns, which are presented in selected areas on Figure~\ref{fig2}.

Figure~\ref{fig3} shows the zoom of the white box on the upper panel of Figure~\ref{fig2}. The clear quasi-periodic pattern starts at 14:00 UT, November 07, when the tornado actually formed, and persists for 7 hr. The period and amplitude of the periodic displacement are gradually increasing with time, but the pattern disappears later. The period is initially about 40 minute, but increases up to 100 minute at the end.

Figure~\ref{fig4} (upper panel) shows the zoom of the white box on the lower panel of Figure~\ref{fig2}. The five panels of the figure reveal the dynamics of the tornado at the heights corresponding to five upper cuts shown in Figure~\ref{fig1}. The clear quasi-periodic transverse displacement of the tornado axis starts at 21:00 UT, November 7 and continues until 02:00 UT, November 8 at each height. The mean period of the displacement is about 50 minute.

There are three possible mechanisms, which may explain the quasi-periodic displacement of the prominence tornado axis: MHD kink oscillations of the tornado, the rotation of two tornado threads around the common axis and the rise of the twisted magnetic tornado during its expansion phase.

Real transverse displacement of the tornado axis is a natural explanation of the observed oscillations. Figure~\ref{fig4} shows that the displacements at different heights are in phase, which indicates a standing pattern of oscillations in this case. There is also evidence that the amplitude of the oscillations increases with height, while no significant oscillation is seen near the footpoint. This probably shows that the whole tornado is oscillating in the fundamental harmonic of MHD kink waves with one fixed end in the photosphere and the second, open end at the top. In this case, the wavelength of fundamental harmonic is four times longer than the actual length of the tornado leading to the value  4$\cdot$50 Mm=200 Mm. Then, one can estimate the kink speed as $c_k \sim$ 70 km s$^{-1}$ inside the prominence tornado. Consequently, the Alfv\'en speed inside the prominence tornado is $V_A\sim c_k/\sqrt{2}\approx$ 50 km s$^{-1}$. Using a typical density for prominence plasma of 5 $\times$ 10$^{-14}$ g cm$^{-3}$ \citep{Labrosse2010}, we can estimate the magnetic field strength in the tornado as $\sim$ 4 G, which corresponds to previous estimations of the magnetic field strength in quiescent prominences \citep{Mackay2010}.

On the other hand, the rotation of two (or more) tornado threads around a common axis may also lead to the apparent transverse displacement. Indeed, space-time plots of Fig.~\ref{fig4} show that the tornado sometime appears as a double structure, which probably is caused due to the rotation \citep{Su2012,Su2014}. For example, the tornado splits into two threads during 60-70 minute (see last three cuts from above) and 120-150 minute (second and third cuts from above) on Fig.~\ref{fig4}, but the threads reunite again. In this case, the rotation period of the threads can be estimated as $\sim$ 90 minute, which leads to the rotational speed of the threads as $\sim 6$ km s$^{-1}$. In principle, both processes do not contradict each other, therefore we may suggest that the tornado is rotating and oscillating at the same time.

The upward expansion of a twisted tornado may also cause an apparent transverse displacement of the tornado axis \citep{Panesar2013,Panasenco2014}. The second oscillation pattern can be hardly explained by this mechanism, because the tornado was not rising during the interval of oscillation. However, this mechanism can be responsible for the first oscillation pattern (see Fig.~\ref{fig3}), because the tornado was rising in height during this interval. Indeed, the apparent oscillation was stopped at the same time when the tornado finished its rising phase. The increase of the oscillation period may indicate that the upward speed of the tornado decreased at the final stage.

At 15:00 UT, November 08, the tornado started to become unstable and eventually erupted together with the associated prominence as CME. The LASCO CME catalog \citep{Yashiro2004} reports the first appearance of the CME in the field of view of the C2 coronagraph at 15:36 UT with a mean speed of 720 km s$^{-1}$. The CME appeared with a very narrow width of 24$^\circ$ in the southwest limb at a central position angle of 160$^\circ$. The higher temperature channels of AIA show a cavity structure above the tornado. Figure~\ref{fig5} shows the time evolution of the cavity in composite images, which are produced from AIA 171, 211 and 193 {\AA} filters. The cavity starts expansion at $\sim$ 13:00 UT and continuously rises above the solar limb. This process extends for an hour and the tornado gradually follows the cavity. Expansion of coronal cavities have been also reported by various authors \citep{li2012,Panesar2013} and they estimated the expansion speed  as $\sim$ 2.5 km s$^{-1}$. The instability of cavity prominences can be explained by the magnetic breakout model \citep{Antiochos1999,Aulanier2000,Maia2003,Shen2012}. Fig~\ref{fig6} shows a schematic picture of the model. We suppose that the cavity (and the prominence tornado) is located between two coronal loop systems. Indeed, Figure~\ref{fig5} clearly shows loop system on left-hand side of the cavity. However, a loop system is not well seen on the other side of the cavity, which could be a result of the different viewing angle. According to the breakout model, magnetic reconnection occurs near a null point, which is formed above the cavity after expansion. The reconnection "opens" the restraining overlying magnetic field lines and the prominence tornado erupts as part of the CME. Accompanied movie clearly shows how the cavity rises upwards and opens "field lines". Careful check of all AIA lines did not reveal any indication of brightening, which could accompany the opening of magnetic field. Therefore, either the brightening was very weak, so we could not detect it in AIA lines or the magnetic reconnection occurred due to the slow Sweet-Parker model, which opened magnetic field lines continuously without rapid energy release and we only could see the final stage of cavity eruption.

After the eruption of the prominence tornado, some material falls down toward the surface. Figure~\ref{fig7} shows the consecutive images in 304 {\AA} line. The falling material probably outlines the remaining magnetic structure, which resembles another loop system on the right-hand side of the cavity. This loop system is not seen in hotter spectral lines, but is important for the breakout model \citep{Antiochos1999}.

\section{Discussion and Conclusion}
We studied the dynamics of a solar prominence tornado in high-cadence image sequences in the \emph{SDO}/AIA 171, 193, 211 and 304 {\AA} channels during 2012 November 7-8. The tornado first appeared at 08:00 UT, 2012 November 07 on the solar southwest limb, persisted over $\sim$ 30 hr and started to erupt together with the associated prominence as a CME at 15:00 UT, November 08. The CME was detected at 15:36 UT on 2012 November 08 according to the LASCO CME catalog.

After its first appearance, the tornado was slowly rising in height from the photosphere toward the corona with a mean speed of $\sim$ 1.5 km s$^{-1}$, which is much smaller than the Alfv\'en and/or sound speeds in the corona and corresponds to the expansion speed estimated by \citet{Panesar2013}.
Time-distance plots show two quasi-periodic patterns in the transverse displacement of the tornado axis during the whole life time of the tornado. The first pattern appeared at the early phase of tornado evolution, when the tornado was rising upwards. It persisted for 7 hr with an oscillation period of 40 minute at the initial stage and 100 minute at the final stage. This quasi-periodic transverse displacement can be caused by the upward motion of a twisted tornado and the increase of period at the later phase may be due to the decrease of the upward expansion speed. The second pattern appeared at the developed stage of the evolution when the tornado already stopped rising, therefore this mechanism can be ruled out in this case. The quasi-periodic transverse displacement of the tornado axis was apparent during 5 hr at different heights with a mean period of 50 minute. The displacements were in phase at all heights showing a slight increase of the amplitude with altitude. Thus, we conclude that the tornado is transversally displaced as a whole with a fixed end at the photosphere and an open end in the corona. The displacements can be either caused by standing kink oscillations in the tornado or by the rotation of two tornado threads around the common axis. If the displacement is caused by kink waves then the Alfv\'en speed in the tornado is estimated as 50 km s$^{-1}$ (considering the oscillation as the fundamental harmonic) using the observed length of the tornado and the oscillatory period. This gives a magnetic field strength of 4 G in the tornado for typical prominence density. Consideration of the first harmonic gives the Alfv\'en speed and the magnetic field strength as 25 km s$^{-1}$ and 2 G, respectively.

On the other hand, if the displacement is caused by the rotation of tornado threads around the common axis then the rotational speed of the threads can be estimated as $\sim 6$ km s$^{-1}$ (note, that this is not a rotation in a classical sense, when the whole tornado is rotating as a solid body). Time series of 171 {\AA} imagery show the apparent rotation of the tornado as a whole body, which could be a "vortical illusion" as suggested by \citet{Panasenco2014}. However, recent spectroscopic observations of \citet{Su2014} showed persistent blue and red Doppler shifts on the two opposite sides of the tornado, evidencing rotational motion of the tornado. Note, that the neutral line between blue and red shifts in spectroscopic observations of \citet{Su2014} shows some evidence of transverse displacement, which could be related to the transverse oscillation of the tornado axis, however it could be also partially due to the jitter effect that was not corrected.

Rotation may split the $m=1$ and $m=-1$ modes of kink waves in tornado, where $m$ is the azimuthal wave number, to have different frequencies. Then the two modes may lead to the spitting of tornado in two oscillatory threads just as it is observed in our case. It would be interesting to study this point theoretically in the future.

The tornado became unstable near 15:00 UT, November 08. The composite movie in 171, 193 and 211 {\AA} lines clearly shows a coronal cavity above the prominence tornado, which expands upwards and finally leads to an erupting CME. We suggest that the magnetic breakout model \citep{Antiochos1999} may explain the observed evolution of prominence/cavity structure (see Fig~\ref{fig5}). Upward expansion of the cavity leads to magnetic reconnection at the null point above the cavity, which "opens" the overlying magnetic field lines allowing the structure to erupt and to develop into a CME. However, we could not see any brightening in AIA spectral lines near the apex of cavity, which may accompany the reconnection. A possible explanation is that the magnetic reconnection occurred due to the Sweet-Parker model, which opened magnetic field lines continuously without a rapid energy release.

If one considers that the second oscillation pattern is caused by standing kink waves then our estimation gives the value of the magnetic field strength in the tornado as 4-5 G. The rotation may twist the magnetic field inside the tornado. However, the azimuthal component can not be much stronger than the axial component as the tube becomes unstable to kink instability after some threshold value. The instability criterion for simple homogeneous static tubes is $B_{\phi}>2B_z$, where $B_{\phi}$ and $B_z$ are azimuthal and axial components of the magnetic field \citep{Lundquist1951,Zaqarashvili2010,Zaqarashvili2014}. Therefore, the azimuthal component of the magnetic field in the tornado can not be more than 10 G. One can speculate that the magnetic field is gradually twisted due to the rotation (if the rotation is function of altitude) until it becomes unstable to kink instability. Then the tornado destabilizes the prominence, which expands upwards and through magnetic breakout model it eruptes and develops to a CME.

In fact, tornadoes could be unstable to kink instability if the apparent transverse oscillation is the result of twisted tube expansion \citep{Panesar2013,Panasenco2014}. Then tornadoes generally may cause instabilities of prominences and thus could be used as precursors for CMEs and consequently for space weather predictions. Therefore, to study the statistical relation of the tornados and unstable prominences is an important task. Initial results show that CMEs can be produced by prominences with tornado-like structures at their footpoints. For example, two other CMEs that occurred on 2012 December 14 at 02:00 UT and 2013 January 22 at 03:48 UT show tornado-like structures near the prominence footpoints. A detailed statistical analysis is under study.

%
%

{\bf Acknowledgements} The work was supported by the Austrian "Fonds zur F\"{o}rderung der Wissenschaftlichen Forschung" (FWF) under project P26181-N27, by FP7-PEOPLE-2010-IRSES-269299 project- SOLSPANET and by Shota Rustaveli Foundation grant DI/14/6-310/12 and DO/233/6-310/13.

\appendix


\clearpage
\begin{figure}
\epsscale{0.43}
\plotone{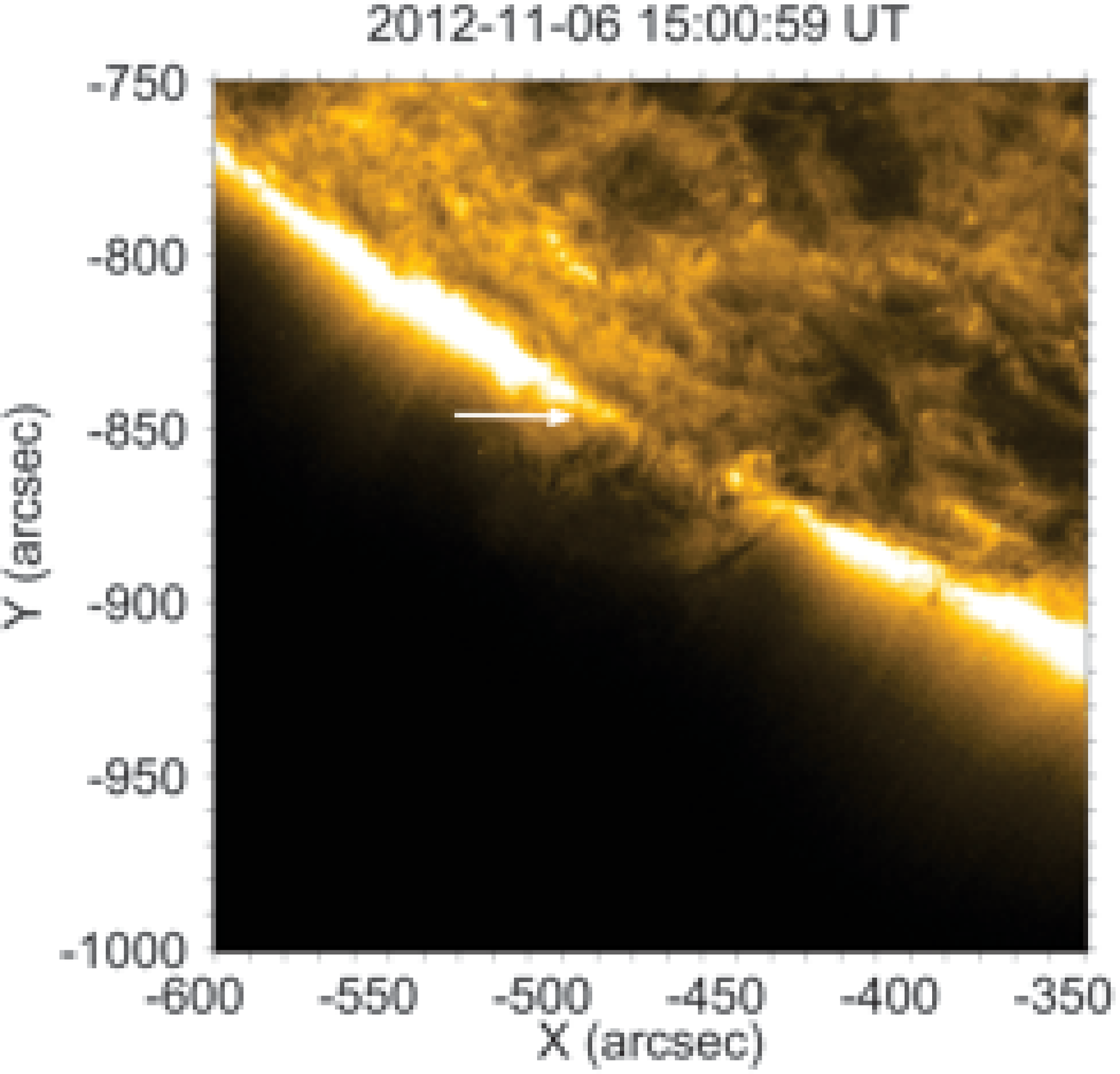}
\plotone{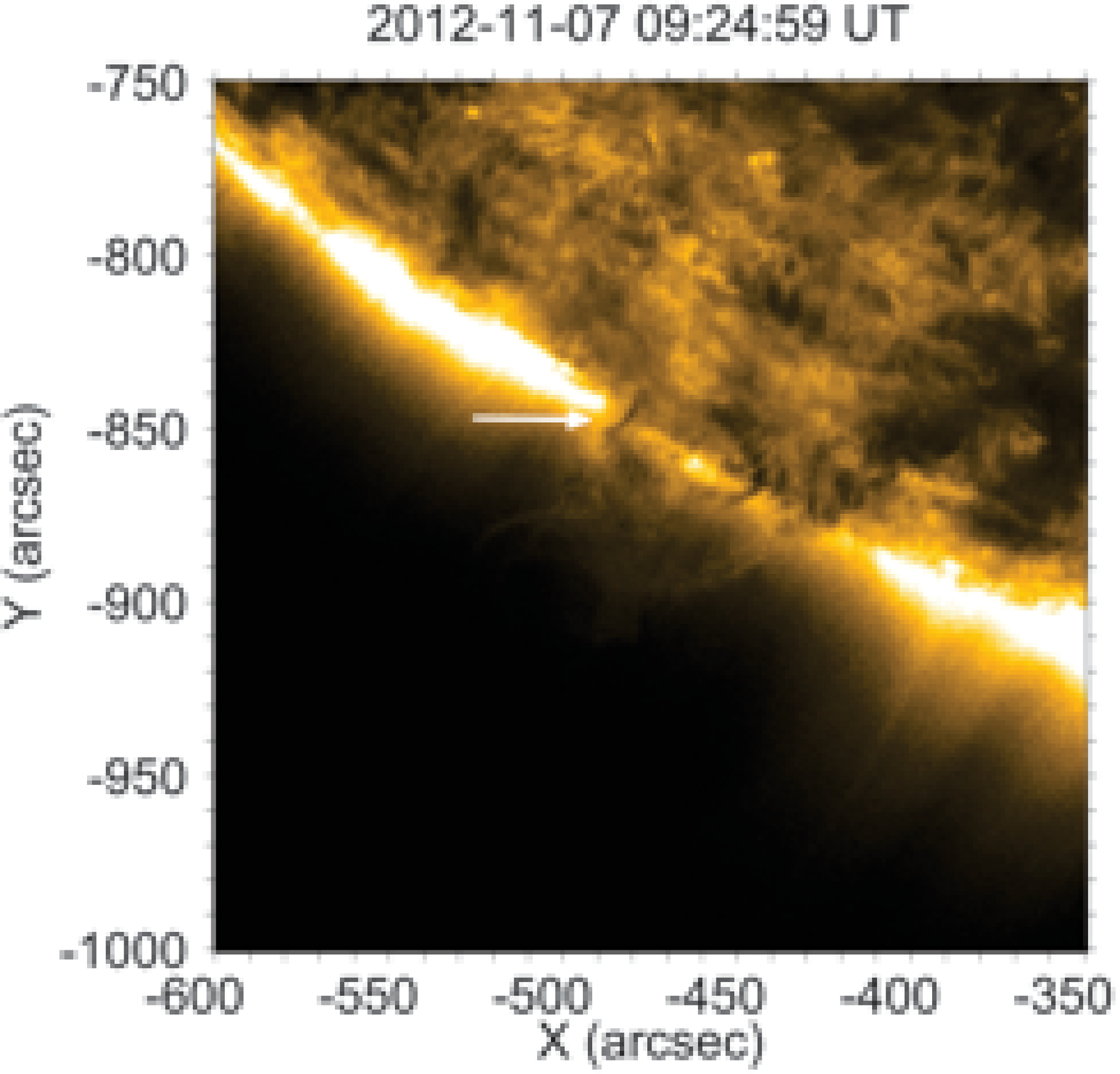}
\plotone{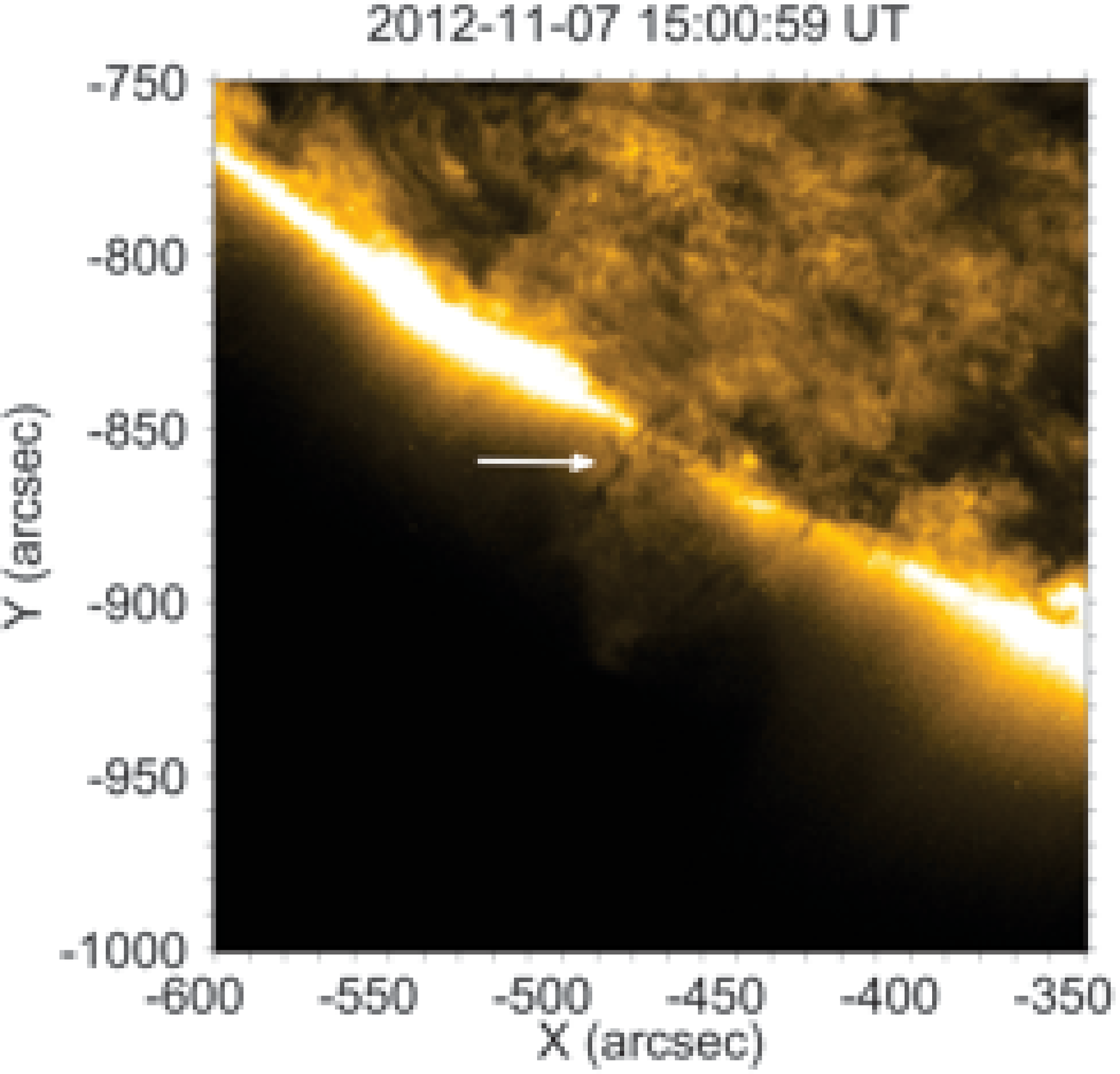}
\plotone{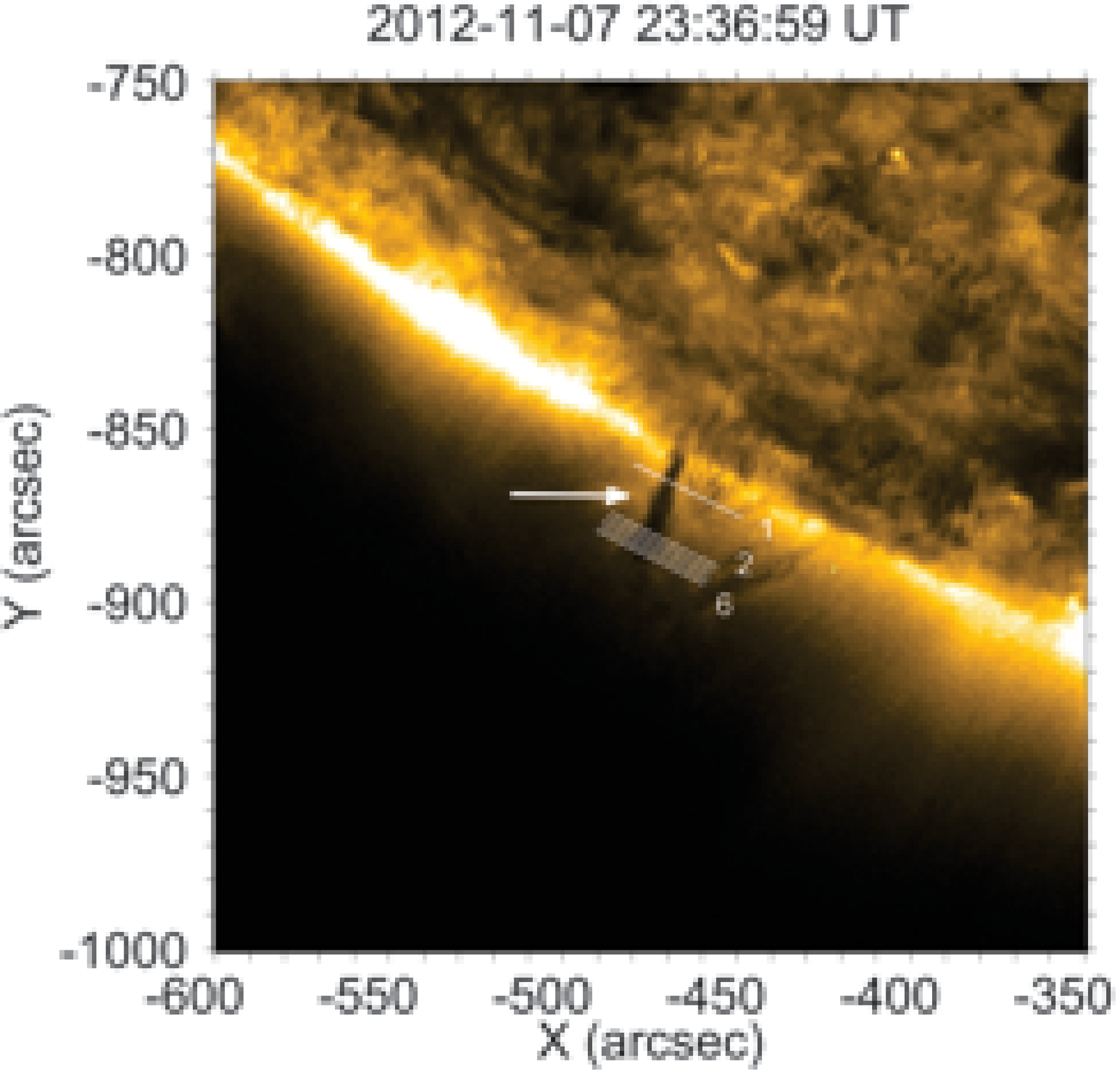}
\plotone{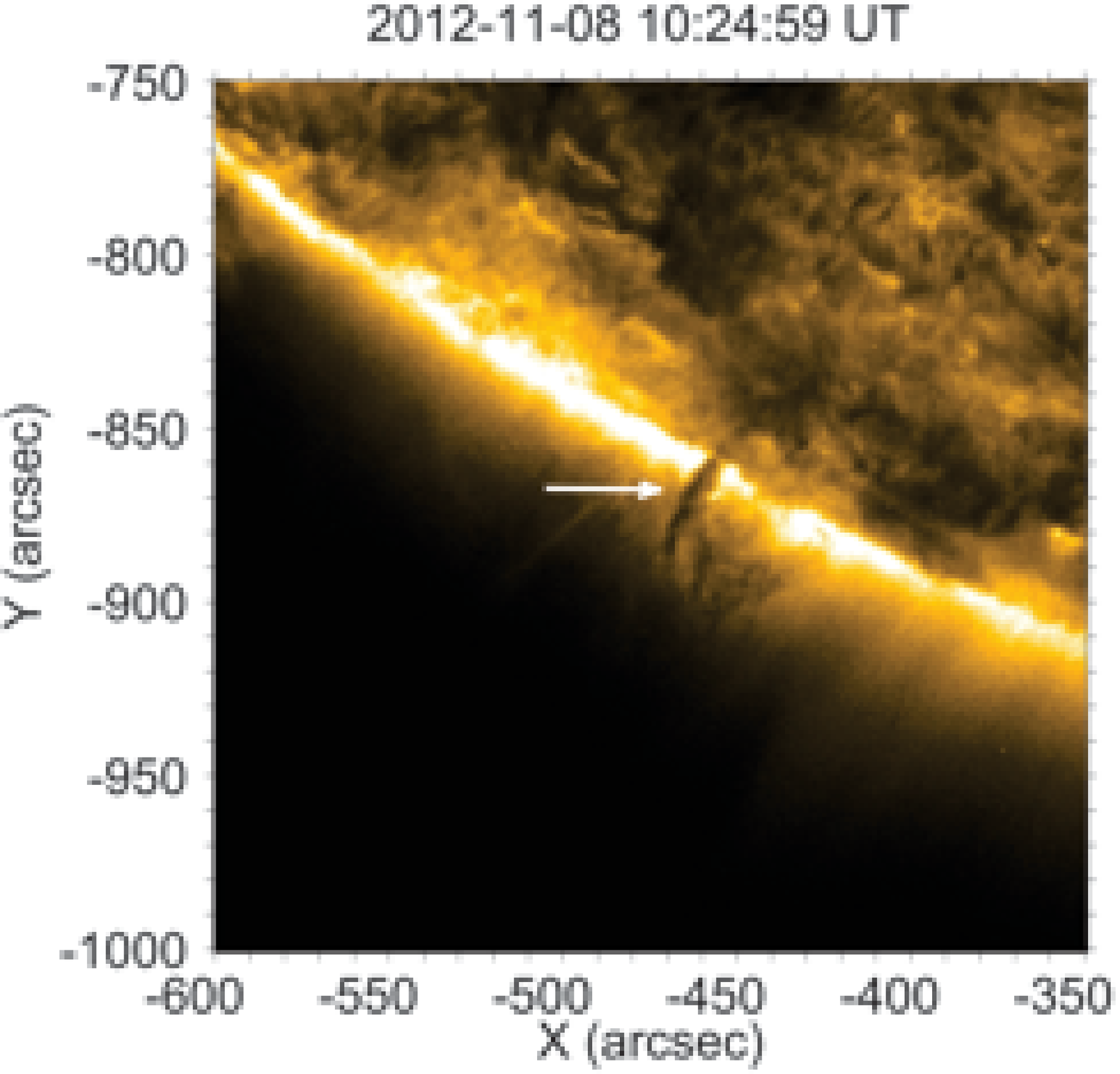}
\plotone{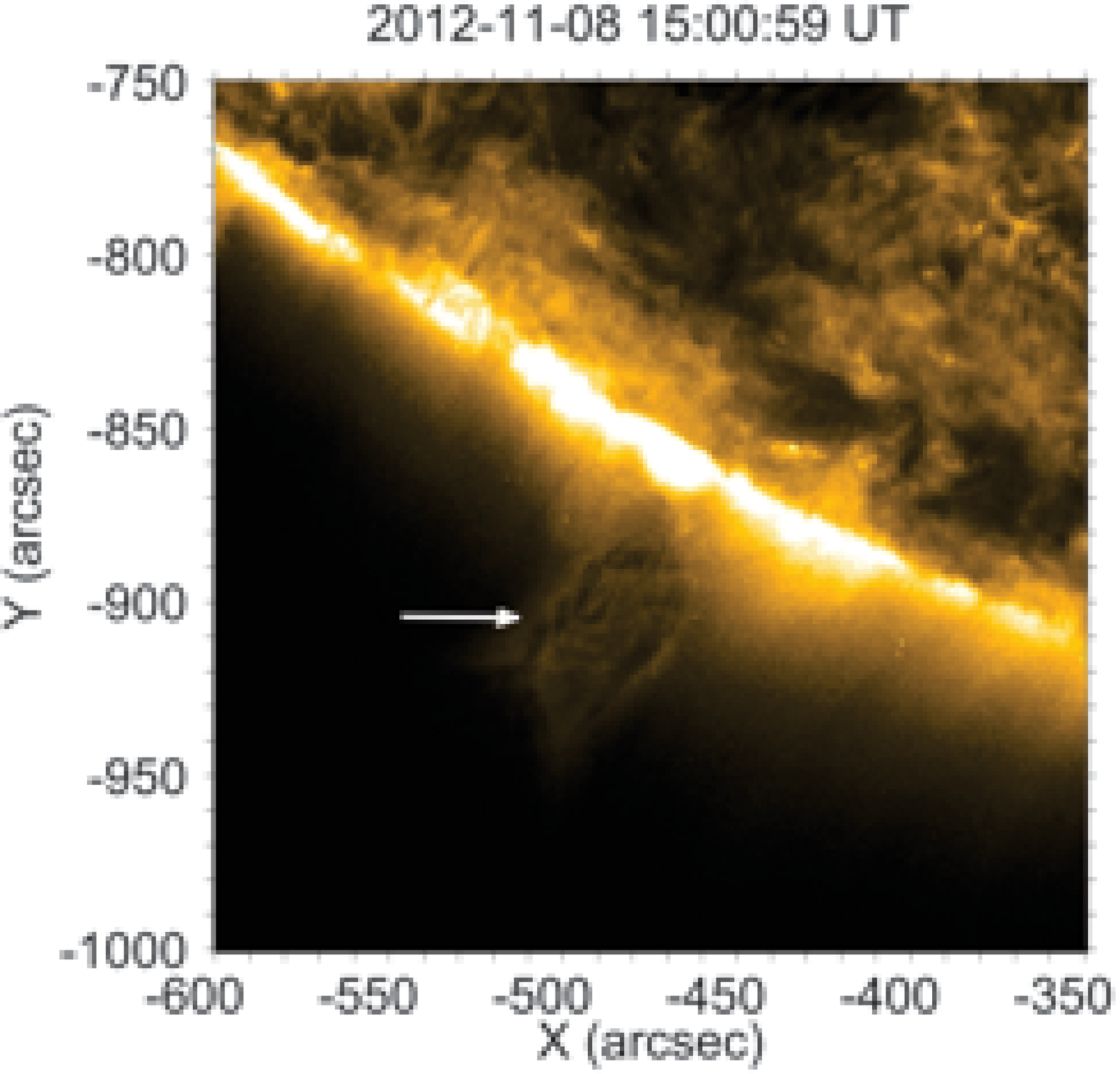}
\caption{Consecutive \emph{SDO}/AIA 171 {\AA} images showing the formation, evolution and eruption of a tornado during 2012 November 07-08. White arrows show the locations of the tornado at different times during its evolution. The solid lines on the middle right panel show the locations of cuts which are used to construct the time-distance plots at different heights shown in Figures 2-4. 
(The corresponding movie is available in the online journal.)
} \label{fig1}
\end{figure}

\clearpage

\begin{figure}
\epsscale{1}
\plotone{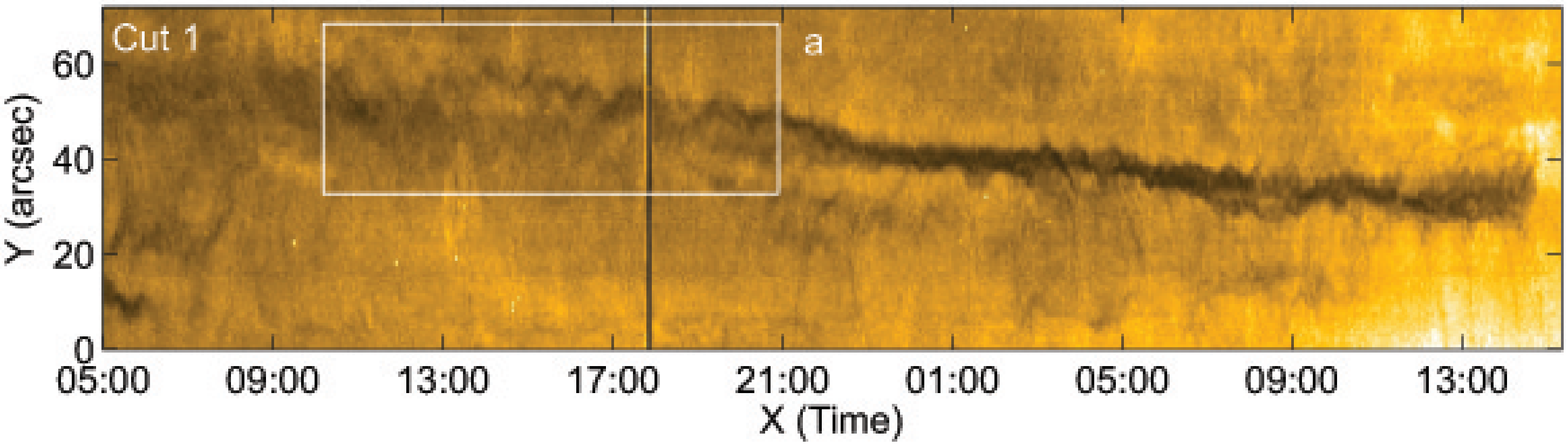}
\plotone{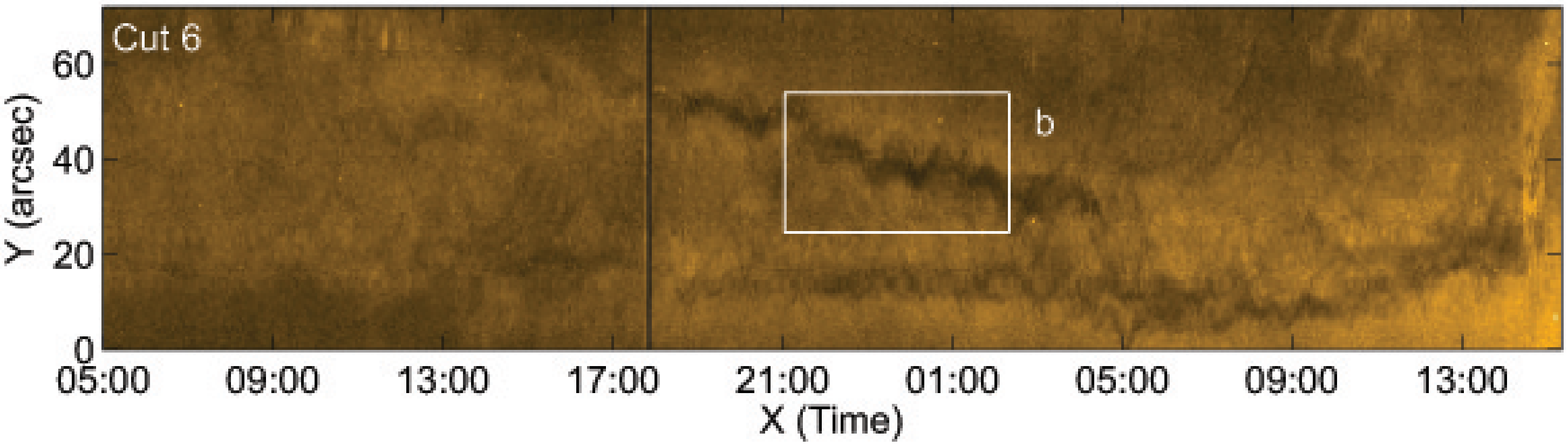}
\caption{Time-distance diagrams corresponding to the cut 1 (upper panel) and cut 6 (lower panel) as shown on the middle right panel of Figure~\ref{fig1}. The start time corresponds to 05:00 UT, 2012 November 07. White boxes labeled a and b show the patterns of quasi-periodic transverse displacements, which are displayed on Figures~\ref{fig3} and ~\ref{fig4}, respectively.
} \label{fig2}
\end{figure}

\begin{figure}
\epsscale{1}
\plotone{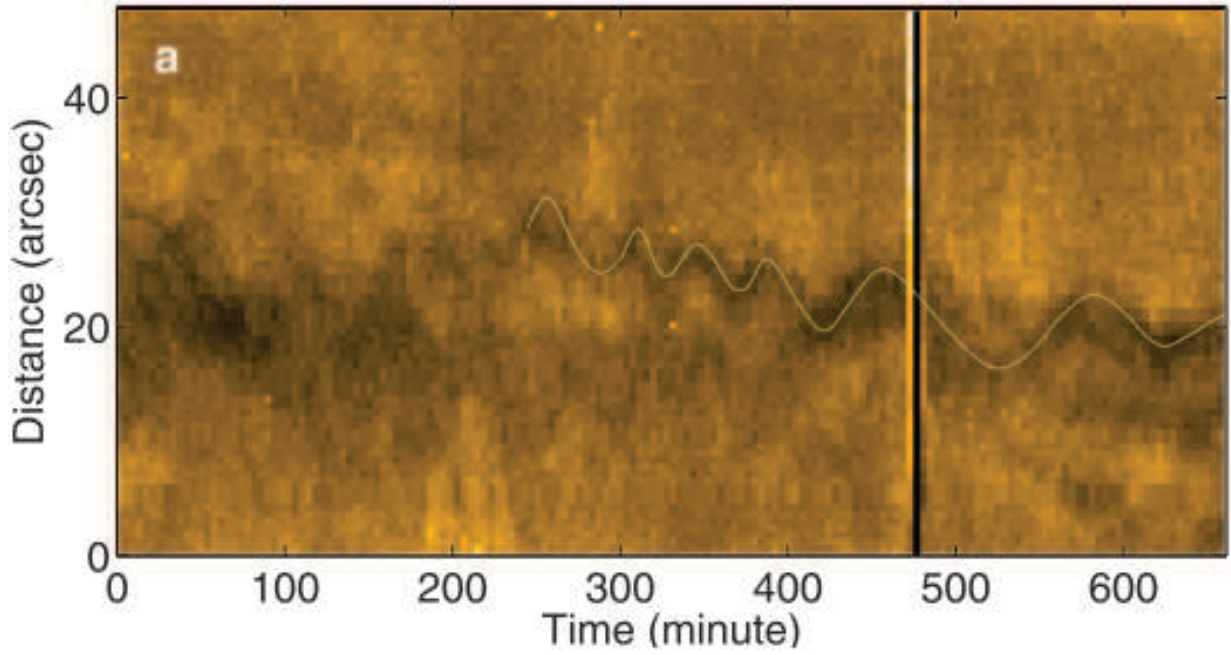}
\caption{Quasi-periodic transverse displacement at cut 1 as shown by the white box a on upper panel of Figure~\ref{fig2}. The start time corresponds to 10:00 UT, 2012 November 07.
} \label{fig3}
\end{figure}

\begin{figure}
\epsscale{0.7}
\plotone{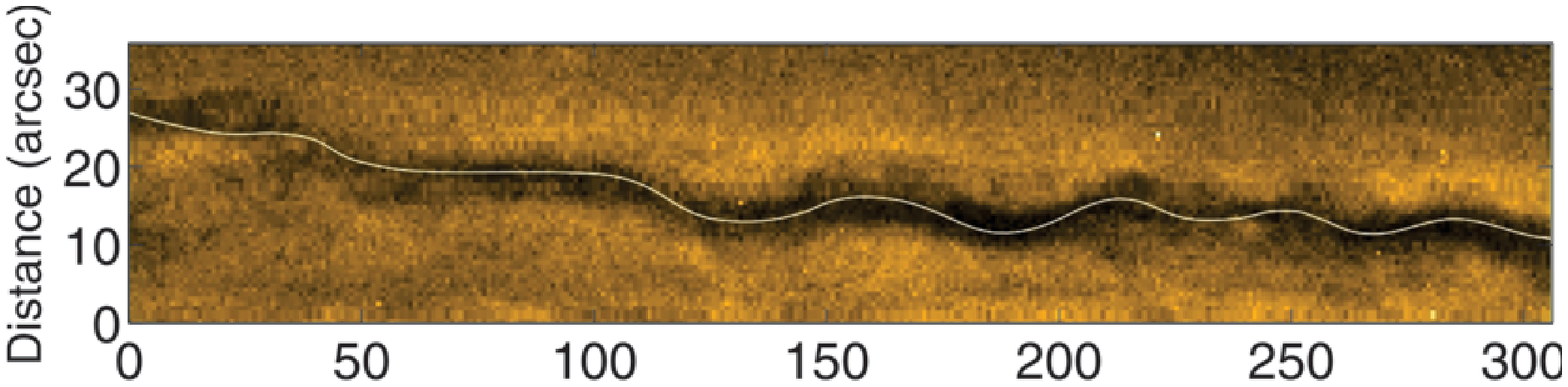}
\plotone{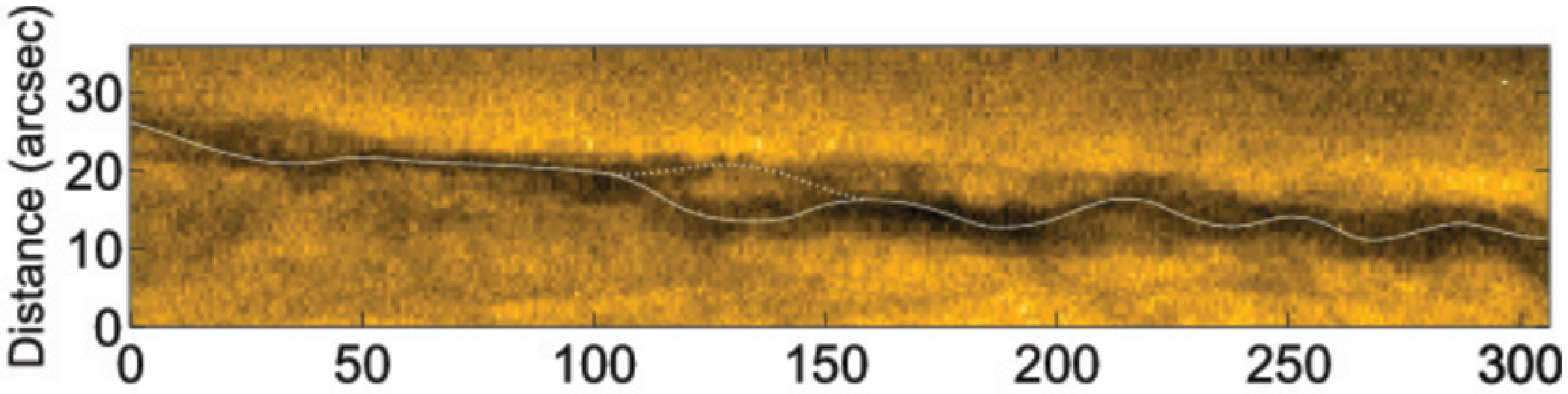}
\plotone{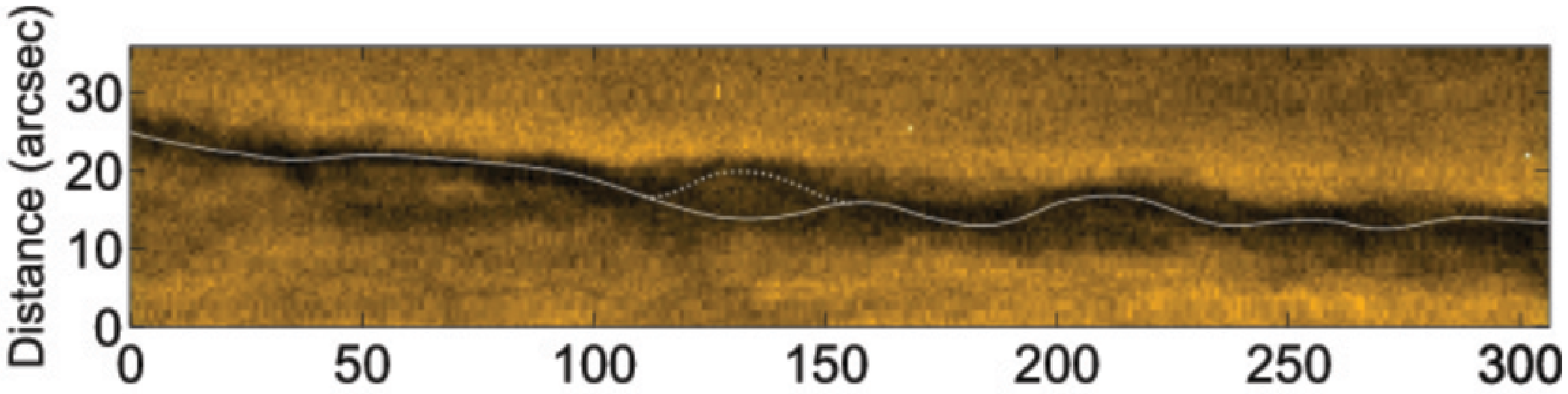}
\plotone{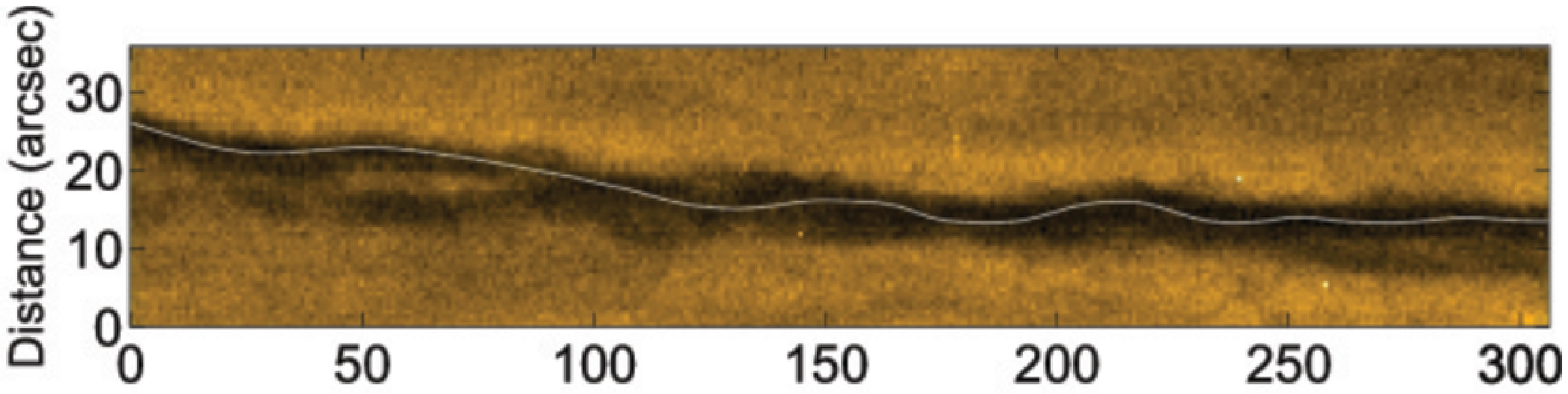}
\plotone{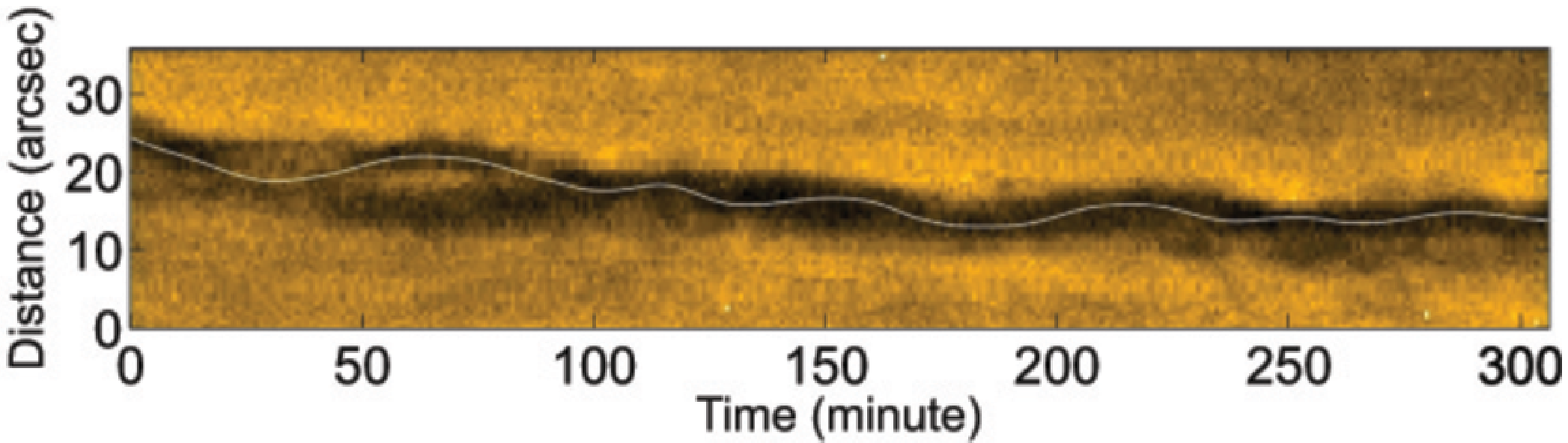}
\caption{Quasi-periodic transverse displacements at five different height levels in the corona corresponding to cuts 2-6 on the middle right panel of Figure~\ref{fig1}. The upper panel displays the dynamics along the cut 6 as shown by the white box b on the lower panel of Figure~\ref{fig2}. The lower panel displays the dynamics along the cut 2. The start time corresponds to 21:00 UT, 2012 November 07.
} \label{fig4}
\end{figure}

\begin{figure}
\epsscale{0.49}
\plotone{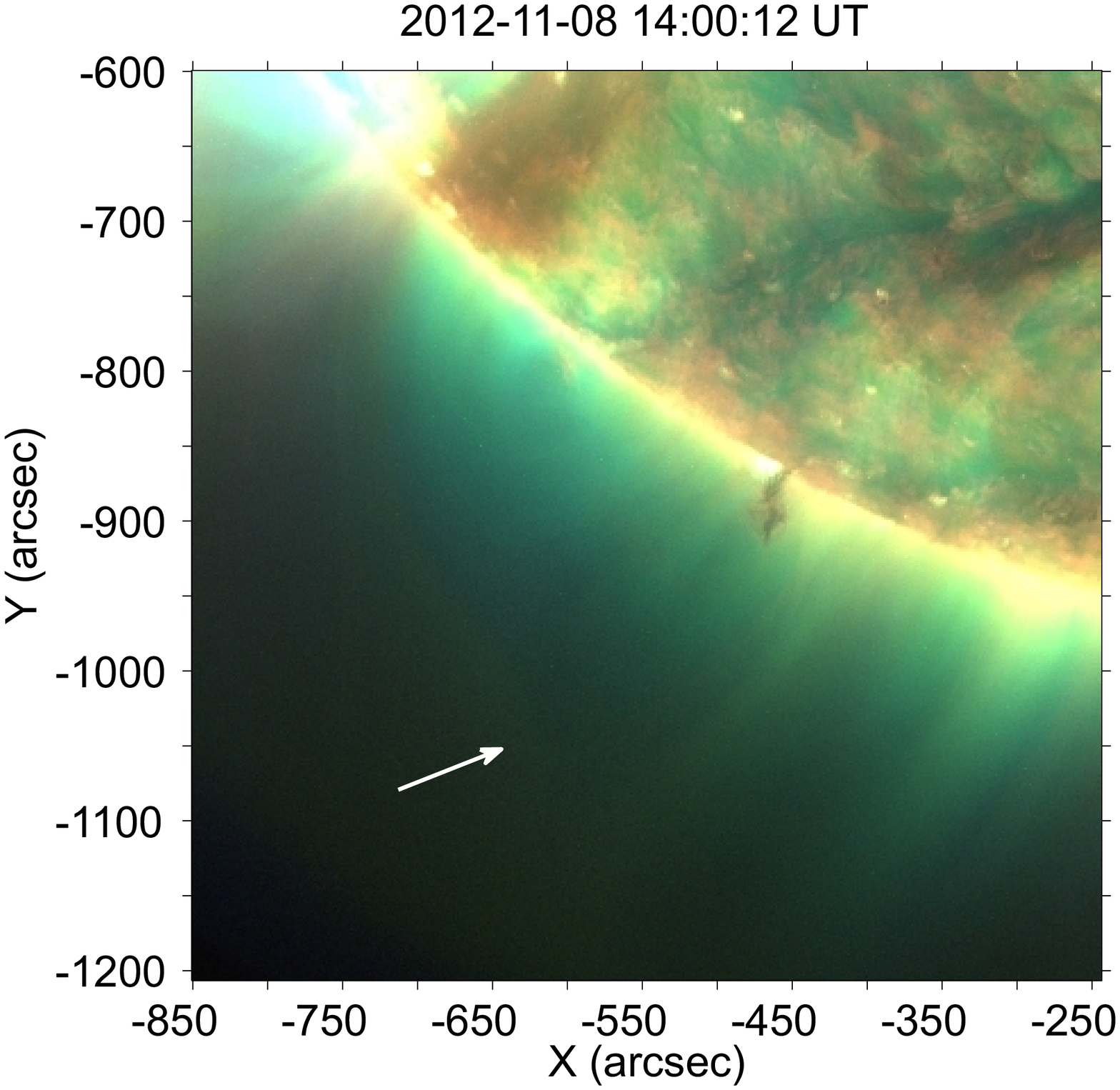}
\plotone{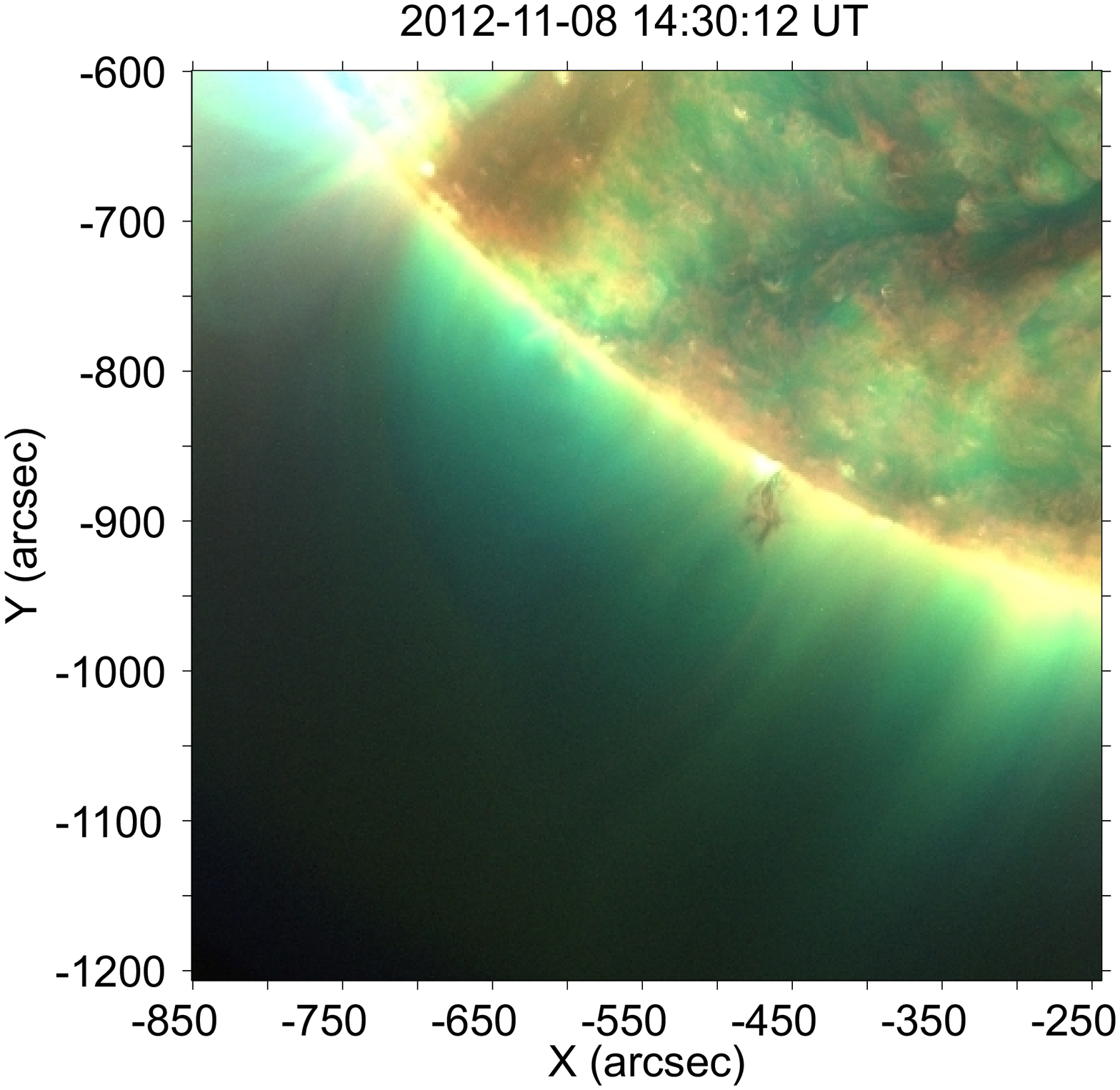}
\plotone{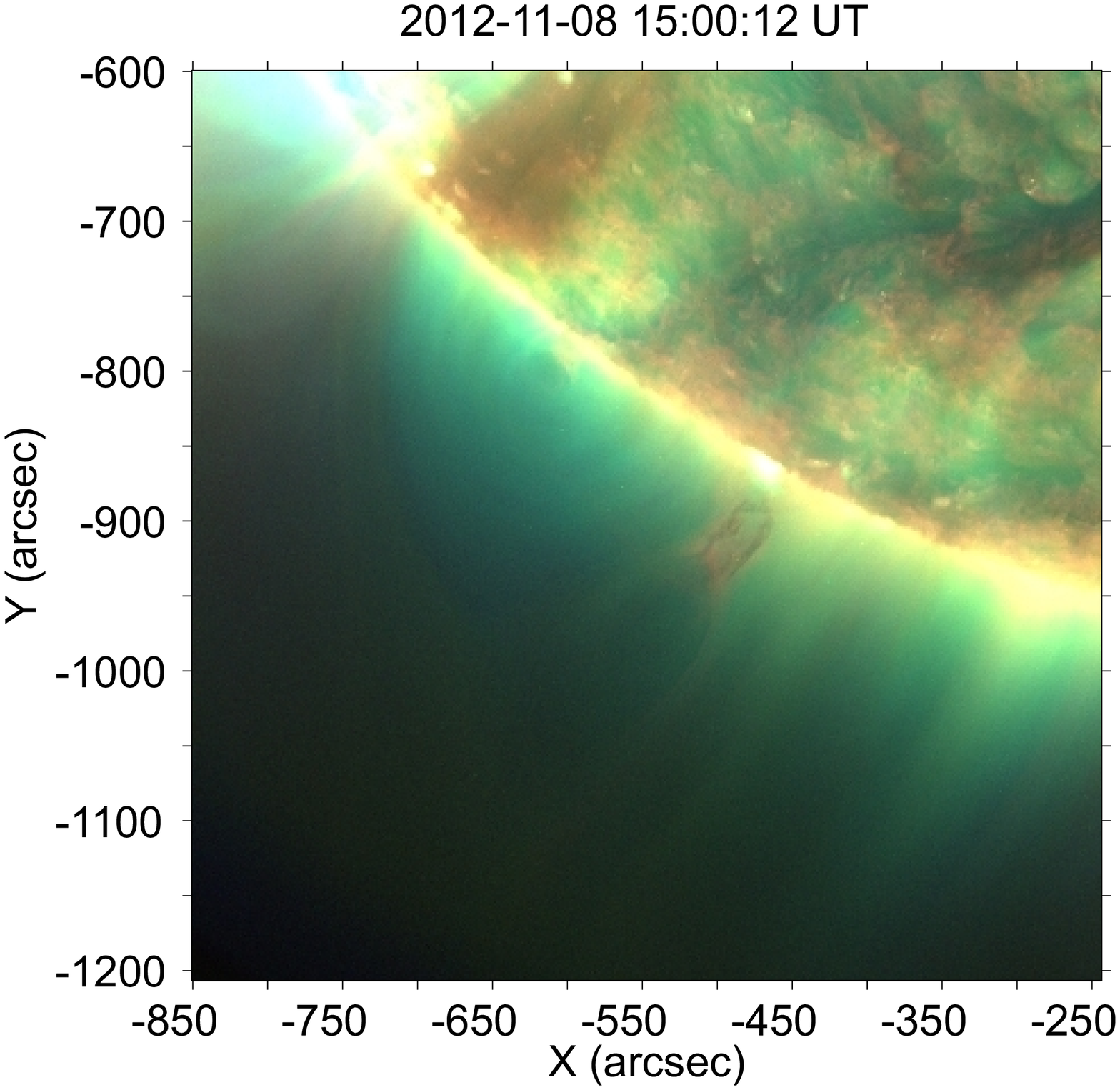}
\caption{Composite images produced from three AIA channels: 171, 211 and 193 {\AA}. A coronal cavity is seen above the tornado, which gradually expands and rises upward. The white arrow indicates the upper boundary of the cavity. The corresponding three-color movie is available in the online journal.
} \label{fig5}
\end{figure}

\begin{figure}
\epsscale{0.49}
\plotone{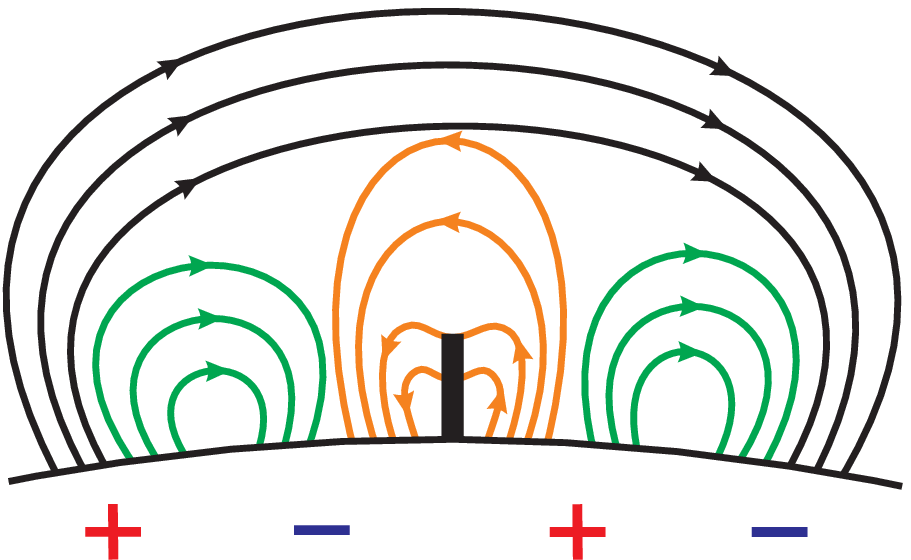}
\plotone{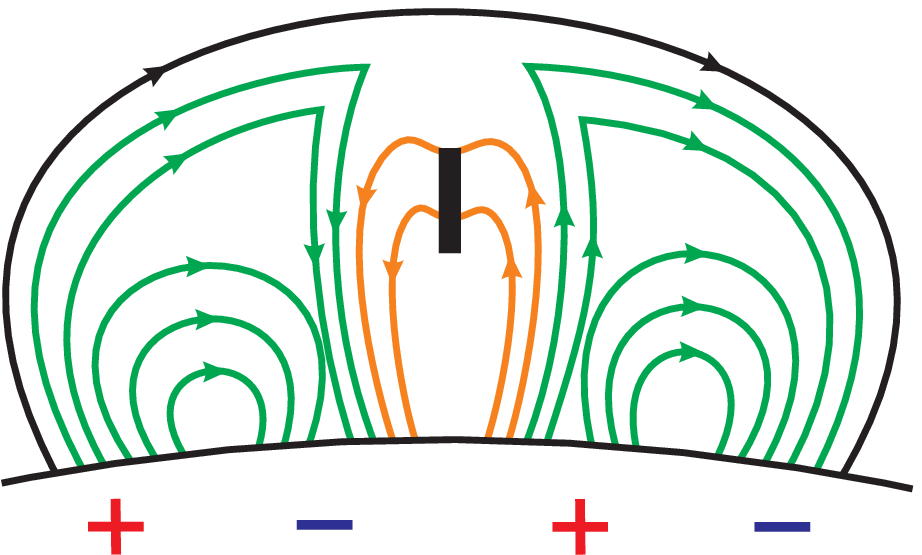}
\plotone{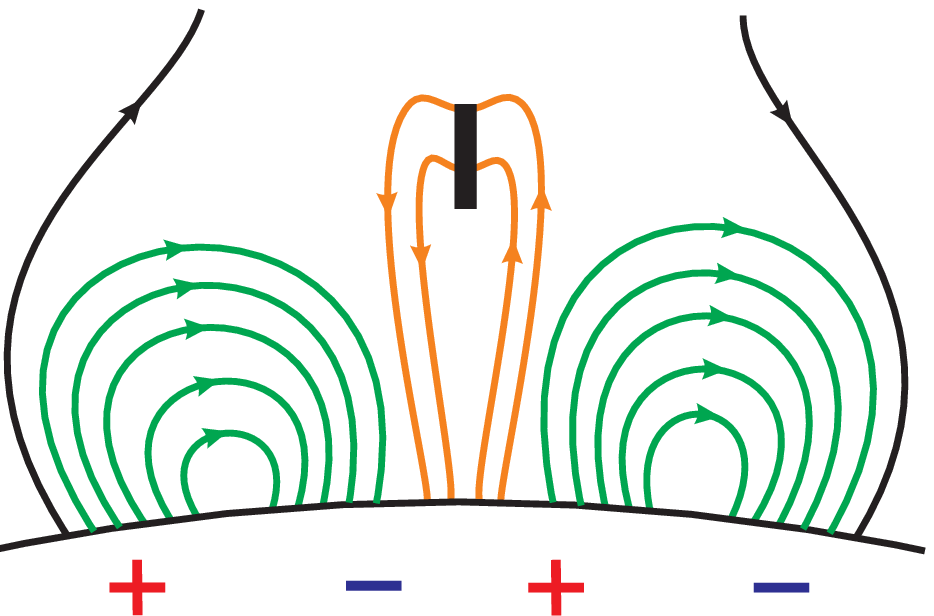}
\caption{Schematic picture of the magnetic breakout model. The $+$ and $-$ signs show the positive and the negative polarities of the magnetic field, respectively. The three stages correspond to the three panels of Figure~\ref{fig5}. The black box represents the tornado.
} \label{fig6}
\end{figure}

\begin{figure}
\epsscale{0.49}
\plotone{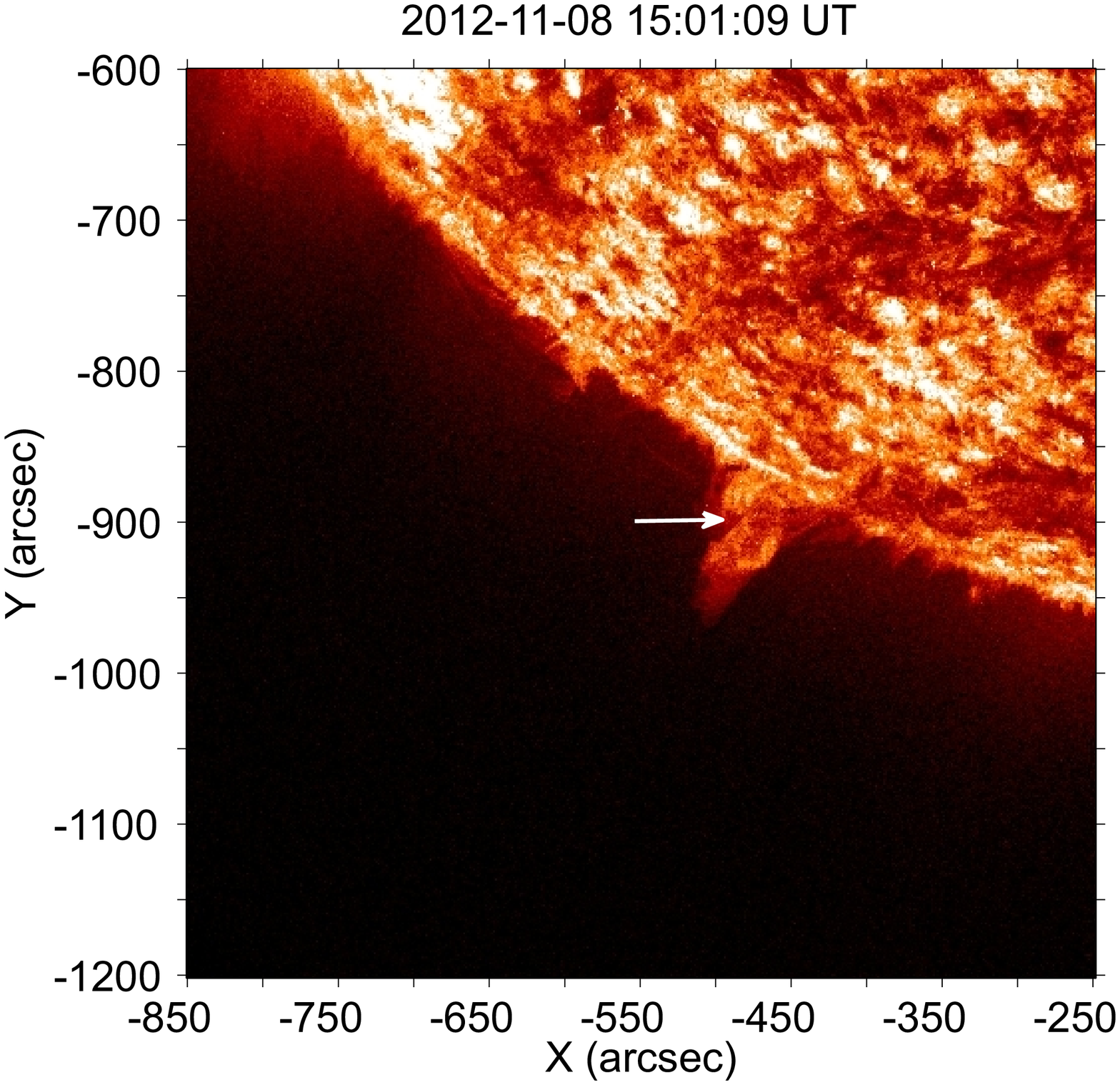}
\plotone{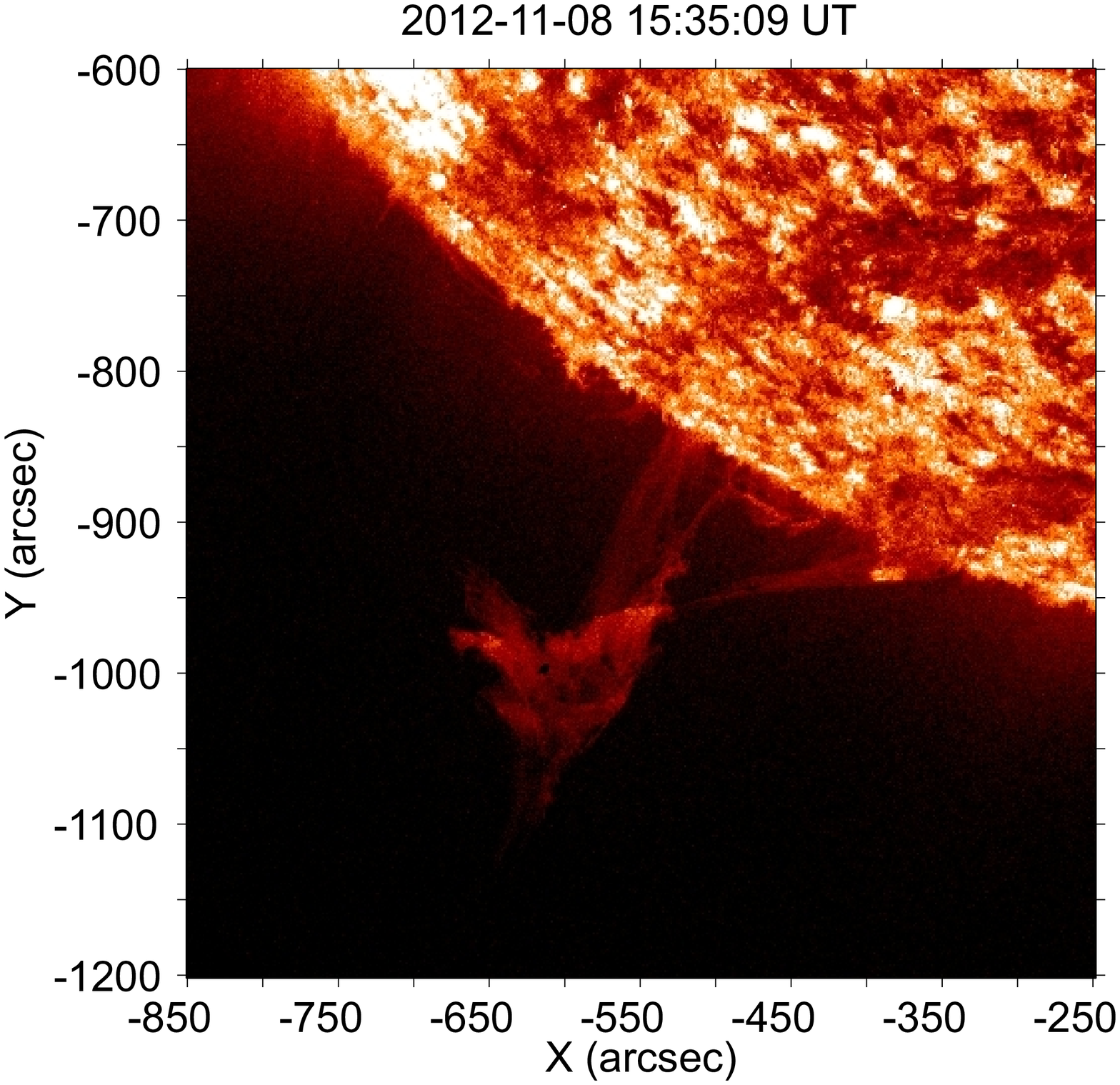}
\plotone{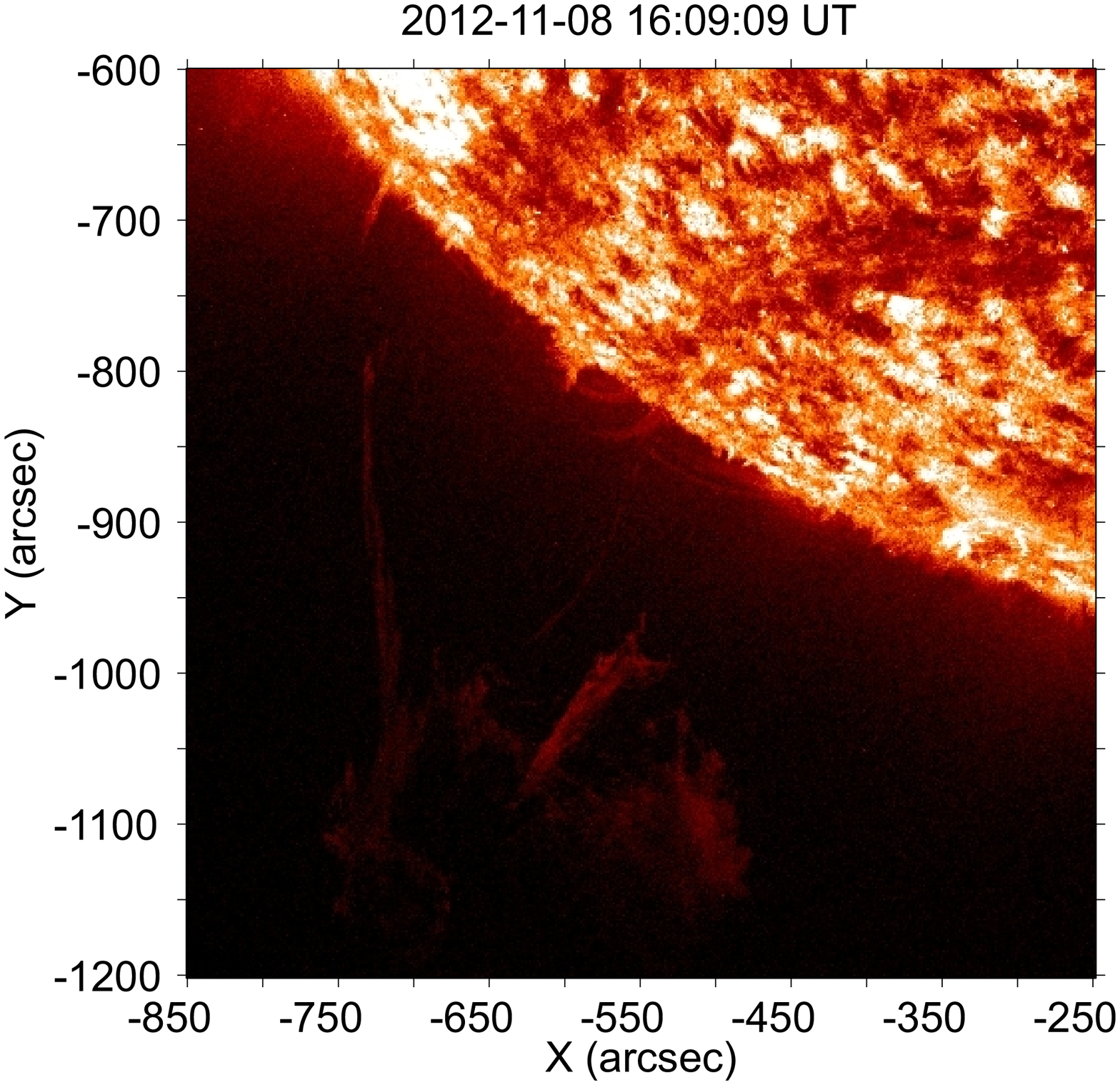}
\caption{Eruptive phase of the prominence observed in the 304 {\AA} filter. The white arrow shows the location, where the tornado appeared in the 171 {\AA} filter. The corresponding movie is available in the online journal.
} \label{fig7}
\end{figure}


\end{document}